\documentclass[%
 aip,
 amsmath,amssymb,
 reprint,%
]{revtex4-1}

\usepackage{graphicx}
\usepackage{dcolumn}
\usepackage{bm}

\usepackage[utf8]{inputenc}
\usepackage[T1]{fontenc}
\usepackage{mathptmx}
\usepackage{etoolbox}
\usepackage[font=footnotesize,labelfont=bf]{caption}
\usepackage[font=footnotesize,labelfont=bf]{subcaption}

\usepackage{color}

\usepackage{amssymb}

\usepackage[figuresright]{rotating}

\makeatletter
\def\@email#1#2{%
 \endgroup
 \patchcmd{\titleblock@produce}
  {\frontmatter@RRAPformat}
  {\frontmatter@RRAPformat{\produce@RRAP{*#1\href{mailto:#2}{#2}}}\frontmatter@RRAPformat}
  {}{}
}%
\makeatother
\begin{document}

\preprint{AIP/123-QED}

\title{How local antipredator response unbalances the rock-paper-scissors model}
\author{J. Menezes}
\email{jmenezes@ect.ufrn.br}
\affiliation{School of Science and Technology, Federal University of Rio Grande do Norte, 59072-970, P.O. Box 1524, Natal, RN, Brazil}
\affiliation{Institute for Biodiversity and EcosystemDynamics, University of Amsterdam, Science Park 904, 1098 XH Amsterdam,
The Netherlands}

\author{S. Batista}%
\affiliation{School of Science and Technology, Federal University of Rio Grande do Norte, 59072-970, P.O. Box 1524, Natal, RN, Brazil}

\author{M. Tenorio}
\affiliation{School of Science and Technology, Federal University of Rio Grande do Norte, 59072-970, P.O. Box 1524, Natal, RN, Brazil}

\author{E. Triaca}
\affiliation{Department of Mechanical Engineering, Federal University of Rio Grande do Norte, Av. Senador Salgado Filho, 300 Lagoa Nova,
59078-970 Natal, RN, Brazil, Brasil}
\author{B. Moura}
\affiliation{Department of Biomedical Engineering, Federal University of Rio Grande do Norte, Av. Senador Salgado Filho 300, Lagoa Nova,
59078-970, Natal, RN, Brazil}
\affiliation{Edmond and Lily Safra International Neuroscience Institute, Santos Dumont Institute\\
Av Santos Dumont, 1560, 59280-000, Macaiba, RN, Brazil}

\date{\today}

\begin{abstract}
Antipredator behaviour is a self-preservation strategy present in many biological systems, where individuals join the effort in a collective reaction to avoid being caught by an approaching predator. We study a nonhierarchical tritrophic system, whose predator-prey interactions are described by the rock-paper-scissors game rules. We performe a set of spatial stochastic simulations where organisms of one out of the species can resist predation in a collective strategy. The drop in predation capacity is local, which means that each predator faces a particular opposition depending on the prey group size surrounding it. Considering that the interference in a predator action depends on the prey's physical and cognitive ability, we explore the role of a conditioning factor that indicates the fraction of the species apt to perform the antipredator strategy. Because of the local unbalancing of the cyclic predator-prey interactions, 
departed spatial domains mainly occupied by a single species emerge.
Unlike the rock-paper-scissors model with a weak species because a nonlocal reason, our findings show that if the predation probability of one species is reduced because individuals face local antipredator response, the species does not predominate.
Instead, the local unbalancing of the rock-paper-scissors model results in the prevalence of the weak species' prey.
Finally, the outcomes show that local unevenness may jeopardise biodiversity, with the coexistence being more threatened for high mobility. 
\end{abstract}

\maketitle

\section{Introduction}
\label{sec:int}

Cyclic models of biodiversity describe nonhierarchical predator-prey interactions among species that promote
the richness of ecosystems in nature \cite{ecology,Nature-bio}. The remarkable outcomes from experiments with bacteria \textit{Escherichia coli}, for example, revealed a cyclic dominance among three bacteria strains, successfully described by the spatial rock-paper-scissors game rules \cite{Coli}. However, the experiment revealed that
the cyclic dominance ensures coexistence only if organisms interact locally, leading to arising departed spatial domains \cite{Allelopathy}. Other authors have also found plenty of evidence that spatial segregation of species is crucial to the formation and stability of ecosystems (for example, in systems with lizards and coral reefs \cite{lizards,Extra1,BUCHHOLZ2007401}).

Stochastic simulations of the spatial rock-paper-scissors model have been widely employed to investigate biological systems where individuals interact locally in a cyclic way \cite{Szolnoki_2020, Szolnoki-JRSI-11-0735,MENEZES2022101606,RANGEL2022104689,ref10-0,ref10-1,ref10-2,ref10-3,ref10-4}. There is evidence that predation and mobility interactions can be influenced by evolutionary behaviour, thus impacting population dynamics and coexistence in cyclic models \cite{Moura,doi:10.1021/ja01453a010,Volterra,PhysRevE.78.031906,0295-5075-121-4-48003,PhysRevE.82.066211,weakest,MENEZES2022111903,ref1-1,ref1-2,ref1-3,ref1-4,ref1-5,ref1-11,ref1-12}. 

It has been shown that
the stability of cyclic models is dependent on the strength the species dominate one another \cite{weakest}: if one species is weaker than the others, in terms of predation capacity, this species predominates \cite{uneven,PedroWeak,Weak4,AVELINO2022111738}. However,
local instabilities - primarily dependent on the initial conditions - can lead to the extinction of two species; in this case, the weaker species is more likely to survive \cite{weakest}. 

The main characteristic of the uneven rock-paper-scissors models studied in literature is that the intrinsic organisms' weakness is not caused by local circumstances but resulting from an evolutionary condition or any external cause. For example, all organisms of one out of the species are affected by a disease outbreak making them less efficient to catch prey, independent of their spatial position \cite{PedroWeak,weakest} . But, in many biological systems, organisms face resistance due to a collective prey self-protection strategy \cite{ContraAtacck2,strategy3}. It has been reported that the effect of the antipredator behaviour is a drop in predation probability, which depends on the prey group size surrounding the predator \cite{manyeyes,dilution1, dilution2}. This means that each predator may be affected differently according to its neighbourhood.
Furthermore, the success of the local antipredator response depends on the organisms' physical and cognitive abilities to detect a nearby enemy and the strength of the self-preservation tactic\cite{Cost2,LizardB1,detection}.

In this work, we study a cyclic nonhierarchical tritrophic system whose predator-prey interactions are unbalanced by local antipredator response performed by organisms of one out of the species. 
Considering that the antipredator response diminishes the organism's predation capacity,  we aim to answer the question of whether a locally weakened species predominates as occurs in the uneven rock-paper-scissors model widely studied, where other reasons than local weaken one out of the species. 

We introduce a conditioning factor parameter that indicates the fraction of individuals apt to join the collective tactic, meaning the percentage of organisms with the necessary physical and cognitive ability to learn and properly execute the antipredator strategy. Furthermore, we assume a maximal distance an organism can influence a predator attack and the strength of the antipredator reaction. 
Our goal is to comprehend how the local antipredator response unbalances the pattern formation and species densities. For this purpose, we follow a numerical implementation recently presented for local antipredator response in rock-paper-scissors models \cite{Anti1,anti2}. 

We explore the emergence of spatial patterns in regions where the cyclic model is locally unbalanced. Besides discovering which species predominates in a locally unbalanced cyclic model, we also focus on the effects of the local unevenness in jeopardising biodiversity, exploring the coexistence probability for a range of mobility probabilities.

\section{The Stochastic Model}

We study a cyclic nonhierarchical system composed of $3$ species, whose predator-prey interactions are described by the rock-paper-scissors game rules. In our model, organisms of one out of the species can react to local predation threats, joining efforts with conspecifics to oppose predator's attacks. Each predator faces a particular antipredator resistance: the larger the prey group surrounding it, the lower the chances of successful predation.

\begin{figure}
\centering
\includegraphics[width=46mm]{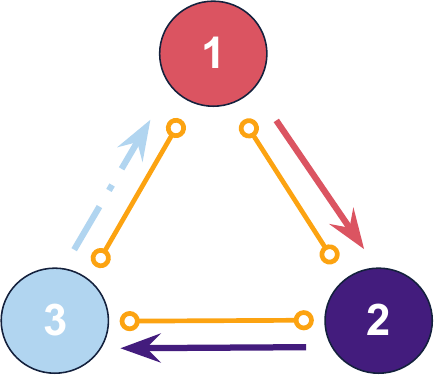}
\caption{Illustration of the cyclic predator-prey interactions in the locally unbalanced rock-paper-scissors model. Red, purple, and light blue arrows represent the dominance of organisms of species $1$, $2$, and $3$, respectively. The dashed arrow illustrates the local reduction in predator capacity of organisms species $3$ by the antipredator behaviour of organisms of species $1$. Orange 
bars indicate that mobility interactions among organisms of every species occurs with same probability.}
\label{fig1}
\end{figure}

Our numerical implementation follows a standard algorithm widely employed in studies of spatial biological systems \cite{anti2,Reichenbach-N-448-1046,PhysRevE.89.042710,Bazeia_2017}. The dynamics of individuals' spatial organisation are simulated in square lattices with periodic boundary conditions, which means that the grid topology is a torus surface. 
 We assumed the Lotka-Volterra numerical implementation, with a conservation law for the total number of individuals \cite{Volterra}; the total number of individuals is $\mathcal{N}$, the total number of grid points.
Figure~\ref{fig1} illustrates the main rules of our simulations, with red, purple, and light blue representing species $1$, $2$, and $3$, respectively. The arrows show the cyclic dominance of the predator-prey interactions: organisms of species $i$ consume individuals of species $i+1$, with $i=1,2,3$, with the cyclic identification $i=i\pm3\,\beta$, where $\beta$ is an integer. The dotted light blue arrow indicates that the predation probability of organisms of species $3$ is locally reduced because of the antipredator behaviour of individuals of species $1$. Orange dots illustrate the mobility interactions among organisms of every species.

\begin{figure}[t]
\centering
\includegraphics[width=46mm]{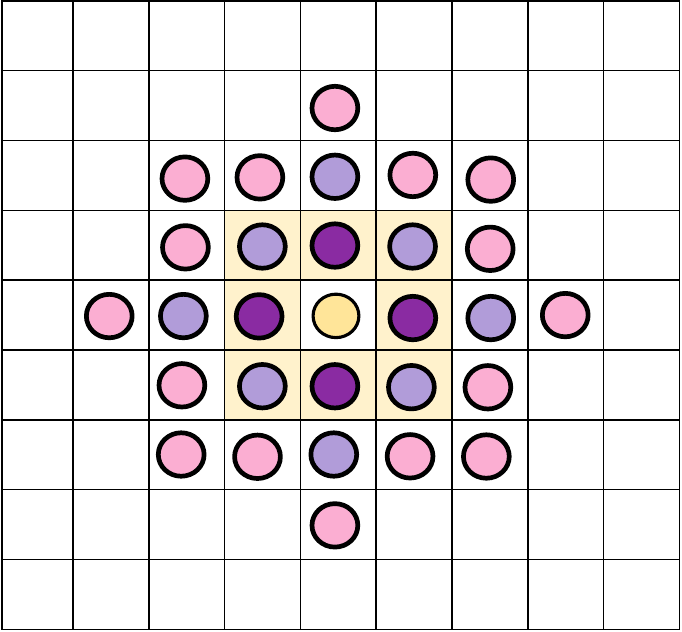}
\caption{Illustration of the Moore neighbourhood and the range of the antipredator response.
An individual positioned at the yellow grid site can interact with one of the eight immediate neighbours (Moore neighbourhood), represented in yellow background.
A predator located at the yellow site faces opposition from the prey group within a range of antipredator response: dark purple dots for $R=1$; dark purple and light purple dots for $R=2$; dark purple, light purple, and pink dots for $R=3$.
}
\label{fig1b}
\end{figure}
\begin{figure*}
\centering
    \begin{subfigure}{.16\textwidth}
        \centering
        \includegraphics[width=29mm]{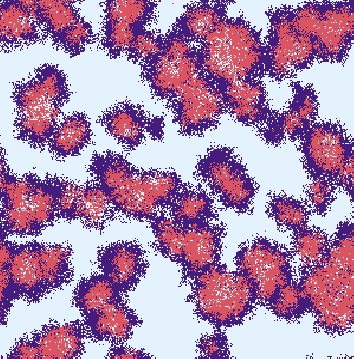}
        \caption{}\label{fig2a}
    \end{subfigure} %
   \begin{subfigure}{.16\textwidth}
        \centering
        \includegraphics[width=29mm]{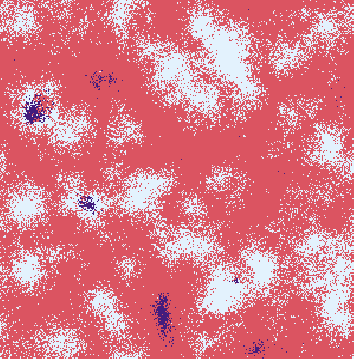}
        \caption{}\label{fig2b}
    \end{subfigure} 
            \begin{subfigure}{.16\textwidth}
        \centering
        \includegraphics[width=29mm]{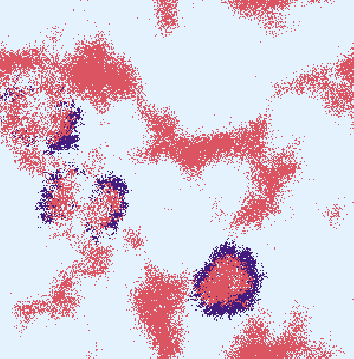}
        \caption{}\label{fig2c}
    \end{subfigure} 
           \begin{subfigure}{.16\textwidth}
        \centering
        \includegraphics[width=29mm]{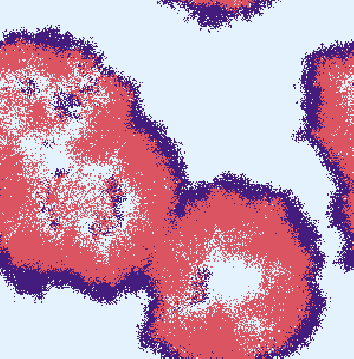}
        \caption{}\label{fig2d}
    \end{subfigure} 
   \begin{subfigure}{.16\textwidth}
        \centering
        \includegraphics[width=29mm]{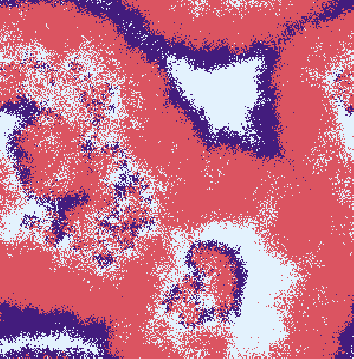}
        \caption{}\label{fig2e}
            \end{subfigure}
   \begin{subfigure}{.16\textwidth}
        \centering
        \includegraphics[width=29mm]{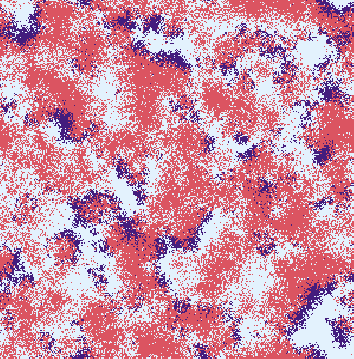}
        \caption{}\label{fig2f}
    \end{subfigure} \\
       \begin{subfigure}{.16\textwidth}
        \centering
        \includegraphics[width=29mm]{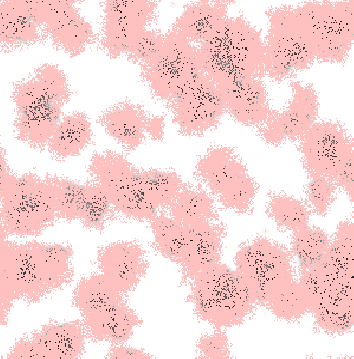}
        \caption{}\label{fig2g}
    \end{subfigure}
          \begin{subfigure}{.16\textwidth}
        \centering
        \includegraphics[width=29mm]{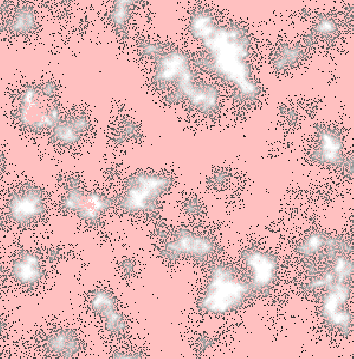}
        \caption{}\label{fig2h}
    \end{subfigure} 
       \begin{subfigure}{.16\textwidth}
        \centering
        \includegraphics[width=29mm]{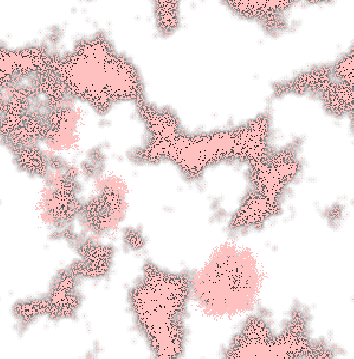}
        \caption{}\label{fig2i}
    \end{subfigure} 
      \begin{subfigure}{.16\textwidth}
        \centering
        \includegraphics[width=29mm]{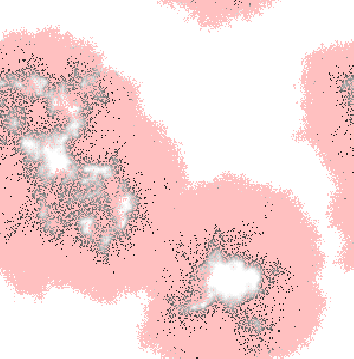}
        \caption{}\label{fig2j}
    \end{subfigure} 
       \begin{subfigure}{.16\textwidth}
        \centering
        \includegraphics[width=29mm]{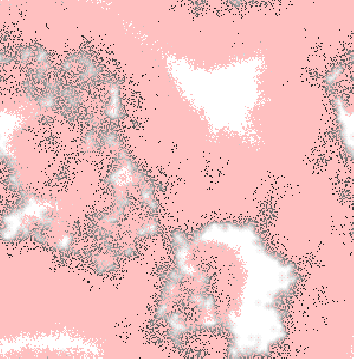}
        \caption{}\label{fig2k}
           \end{subfigure} 
       \begin{subfigure}{.16\textwidth}
        \centering
        \includegraphics[width=29mm]{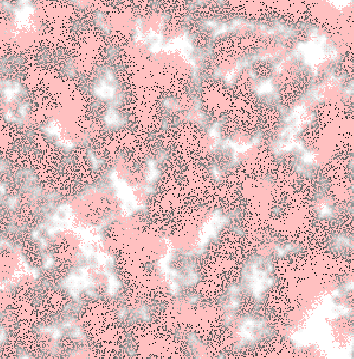}
        \caption{}\label{fig2l}
    \end{subfigure} 
\caption{Snapshots captured from a simulation in a lattice with $300^2$ grid points. The realisation ran until $3000$ generations, for $R=3$, $\kappa=7.5$, $\alpha=1.0$, $p=m=0.5$. Figures a, b, c, d, e, and f show the organisms' spatial distribution after $36$, $60$, $84$, $120$, $144$, and $252$ generations, respectively. The colours follow the scheme in Fig. 1. 
Figures g, h, i, j, k, and l show how predation capacity are spatially distributed in the snapshots of Figs. a, b, c, d, e, and f, respectively. Pink dots represent $\varepsilon_1$ and $\varepsilon_2$, while the shades of grey shows the variation between the minimum (black) and maximum (white) values of $\varepsilon_3$.
}
  \label{fig2}
\end{figure*}

The initial conditions were prepared so that the number of individuals is the same for every species, i.e., $I_i\,=\,\mathcal{N}/3$, with $i=1,2,3$. We allocate each individual at a random grid point. Every time step, one spatial interaction is completed:
\begin{itemize}
\item 
Predation: $ i\ j \to i\ i\,$, with $ j = i+1$. When one predation interaction occurs, an organism of species $i$ (the predator) replaces the grid point filled by the individual of species $i+1$ (the prey).
\item
Mobility: $ i\ \odot \to \odot\ i\,$, where $\odot$ means an individual of any species. When moving, an individual of species $i$ switches positions with another organism of any species.   For example, an organism of species 1 may change spatial position with another individual of species 1, 2, or 3. 

\end{itemize}
We work with the Moore neighbourhood, i.e., individuals interact with one of their eight nearest neighbours, as illustrated by the yellow dot (active individual) and yellow background sites 
(eight possible passive individuals) in Fig.~\ref{fig1b}. The simulation algorithm follows three steps: i) randomly selecting an active individual; ii) raffling one interaction to be executed; iii) drawing one of the eight nearest neighbours to suffer the sorted interaction. Mobility interactions are always implemented because two organisms can switch positions irrespective of their species; however, predation only occurs if the randomly chosen neighbour is the active individual's prey. Therefore, if the randomly chosen interaction is realised, one timestep is counted. Otherwise, the three steps are redone. 
Predation and mobility interactions are chosen with probabilities $p$ and $m$, with $p+m=1$, for every species.
The time necessary to $\mathcal{N}$ timesteps to occur is one generation, our time unit.
The species densities $\rho_i$, with $i=1,2,3$, is defined as the fraction of the grid occupied by individuals of species $i$ at time $t$: $\rho_i\,=\,I_i/\mathcal{N}$.

To explore the local aspects of antipredator behaviour, we define the maximum Euclidean distance at which prey can interfere with the predator action: the radius of the antipredator response $R$,
measured in units of the lattice spacing.
Consequently, the maximum number of individuals participating in a collective reaction against a predator is the number of organisms that fits within a circular area of radius $R$ centred at the predator position, which we define as $\mathcal{G}$. 

Therefore, we define the predation capacity $\varepsilon_i (x,y)$ that represents the probability of a predator of species $i$, located at the spatial position $(x,y)$ in the lattice, consuming a prey present in its immediate neighbourhood. As no antipredator resistance is performed by individuals of species $2$ and $3$, we assume $\varepsilon_1 =\varepsilon_2=1$, independent of the spatial position. On the other hand, each individual of species $3$ has its predation capacity reduced according to the prey group size in the neighbourhood. For a given predator of species $3$, the predation capacity is calculated by means of the Holling type II functional response \cite{holling_1965}:
\begin{equation}
\varepsilon_3\,=\frac{1}{1\,+\,\kappa\,\frac{g}{\mathcal{G}}} \label{ht2}
\end{equation}
where $g$ is the actual group size. This means that the effective predation probability for an organism of species $3$ is given by $p_{eff} =\varepsilon_3\,p$.
The real parameter $\kappa$ is the antipredator strength factor, with $\kappa\geq0$; $\kappa=0$ represents the standard model (the absence of local antipredator response), that is, $\varepsilon_3=1$. In our model, a lonely prey ($g=1$) manages to reduce the effective predation probability to
$\varepsilon_3=1/(1\,+\,\kappa/\mathcal{G})$,
while $\varepsilon_3$ is minimal when $g=\mathcal{G}$, i.e., $\varepsilon_3=1/(1\,+\,\kappa)$. 
In addition, we introduce the conditioning factor $\alpha$, a real parameter, $0\,\leq\,\alpha\,\leq\,1$, representing the percentage of organisms of species $1$ with physical and cognitive ability to perform the collective behavioural antipredator strategy.
\section{Pattern formation}

To study the pattern formation process, we first performed a single simulation in a square lattice with $300^2$ grid points for a timespan of $3000$ generations. All individuals of species $1$ were assumed to be conditioned to participate in the antipredator strategy, $\alpha=1.0$. The radius of antipredator response was set to $R=3$; while $\kappa\,=\,7.5$, $p\,=\,m\,=\,0.5$. We captured $250$ snapshots of the lattice in the first stage of the simulation; then, we used the snapshots to produce the video in https://youtu.be/lF4p7MTwR44. Following the colour scheme in Fig.~\ref{fig1}, organisms of species $1$, $2$, and $3$, are depicted by red, purple, and light blue dots, respectively.
Figures~\ref{fig2a} to \ref{fig2f} show snapshots of the spatial configuration after 
$36$, $60$, $84$, $120$, $144$, and $252$ generations (we have chosen these snapshots to highlight the pattern formation process). 
Because of the local antipredator response of species $1$, the population of species $3$ declines immediately after the simulation begins. Individuals of species $1$ proliferate, consuming almost all organisms of species $2$.
Then, the high density of species $1$ allows the population of species $3$ to grow despite the predator capacity being dropped by the antipredator response.
The cyclic predator-prey interaction makes it possible to  species $2$ to increase; however, Fig.~\ref{fig2a},
reveals that, while advancing over areas dominated by species $3$, individuals of species $2$ are quickly invaded by individuals of species $1$. We observed that the local unevenness introduced by the resistance against predation allows species $1$ to grow faster than the others.

Our outcomes show that after an initial transient creation of expanding domains - responsible for the initial alternate lattice dominance - spatial patterns are formed as shown in Fig. 3f.  This happens because in patches with a low concentration of species $1$, the antipredator response is limited, thus facilitating the multiplication of organisms of species $3$. On the other hand, in areas with many individuals of species $1$, the antipredator response limits the appearance of offsprings of species $3$.

To observe the spatial distribution of organisms with different predation capacities, we calculate $\varepsilon_i$ for each individual during the entire simulation. Figures~\ref{fig2g} to ~\ref{fig2l} shows the results for the snapshots in Figs.~\ref{fig2a}
to ~\ref{fig2f}. Pink dots show the presence of individuals of species $1$ and $2$, whose predation capacity is always maximum, irrespective of the spatial position. We applied a greyscale to distinguish individuals of species $i$ according to their predation capacity: the most affected individuals are depicted in black, while organisms not facing antipredator resistance appear in white; intermediary values of $\varepsilon_3$ are shown in shades of grey. The outcomes show that individuals with  
more decreased predation capacity are 
scattered within regions dominated by species $1$; in contrast, organisms with $\varepsilon_3=1$ are concentrated, forming dense white regions. Therefore, the proportion of individuals not affected by the local antipredator response far surpasses those coping with maximum resistance.

\begin{figure}
\centering
\includegraphics[width=77mm]{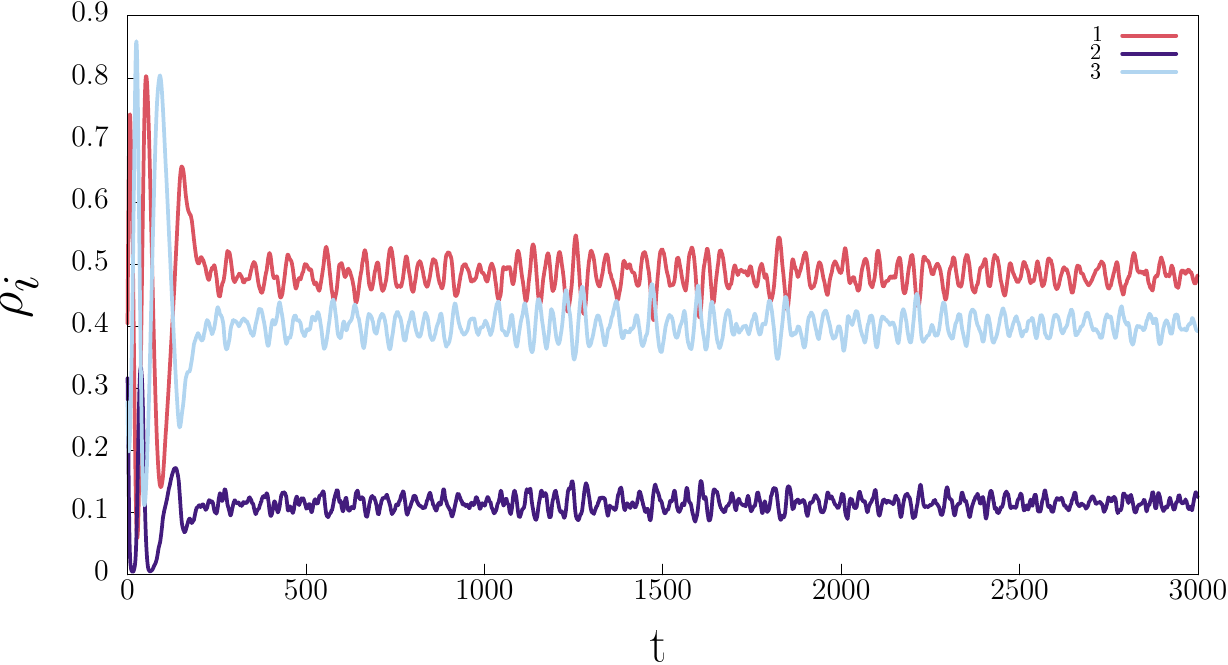}
\caption{Temporal changes of the species densities in the simulation presented in Fig. 3. Red, purple, and light blue lines depict $\rho_i$ for $i=1$, $i=2$, and $i=3$, respectively.}
	\label{fig3}
\end{figure}

\begin{figure}
\centering
\includegraphics[width=77mm]{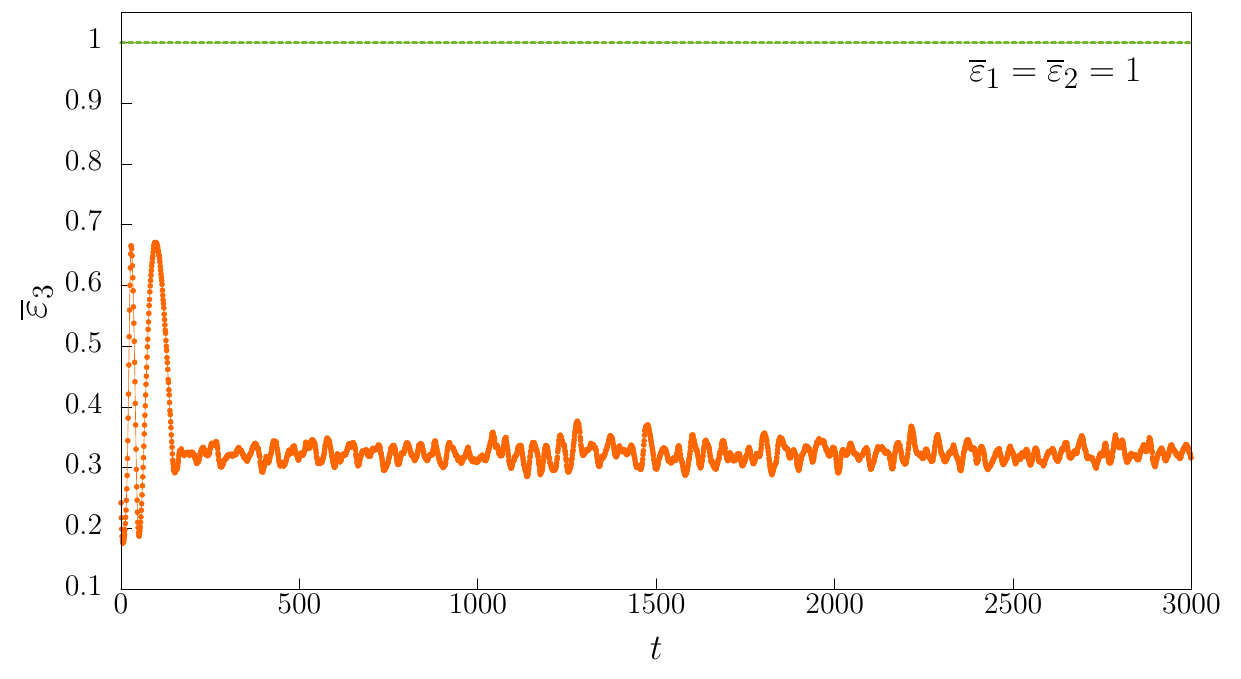}
\caption{Average predation capacity as a function of the time in the simulation presented in Fig. \ref{fig2}. The dashed green lines indicates that $\overline{\varepsilon_1}=\overline{\varepsilon_2}=1$, whereas the orange line shows the dynamics of $\overline{\varepsilon_3}$.}
	\label{fig3b}
\end{figure}

We also calculated the temporal variation of the species densities in the simulation shown in Fig. 3, which is depicted in Fig.~\ref{fig3}. After a short pattern formation period, the average species densities remain constant until the end of the simulation, with $\rho_1>\rho_3>\rho_2$. Additionally, we quantified how the average spatial predation capacity of species $i$, denoted by $\overline{\varepsilon_i}$, changes during the simulation. The orange line in Fig.~\ref{fig3b} depicts the time dependence of $\overline{\varepsilon_3}$, while the green dashed line indicates that $\overline{\varepsilon_1}=\overline{\varepsilon_2}=1$. 
After a rapid variation in the initial stage, $\overline{\varepsilon_3}$ fluctuates around a constant value. Figure~\ref{fig3b} confirms that species $3$ is weaker than the others (in the sense that the average predation capacity is lower). Nevertheless, according to Fig.~\ref{fig3}, here, the weaker species does not preponderate over species $1$ and $2$, as it occurs if the species is weaker because of an intrinsic nonlocal condition \cite{PedroWeak,weakest,uneven}.


\begin{figure*}[t]
\centering
    \begin{subfigure}{.18\textwidth}
        \centering
        \includegraphics[width=32mm]{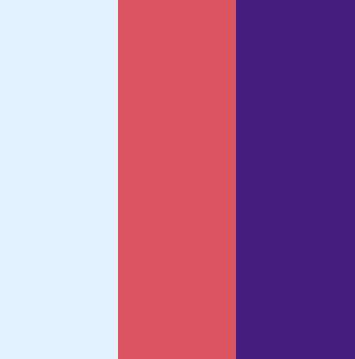}
        \caption{}\label{fig4a}
    \end{subfigure} %
       \begin{subfigure}{.18\textwidth}
        \centering
        \includegraphics[width=32mm]{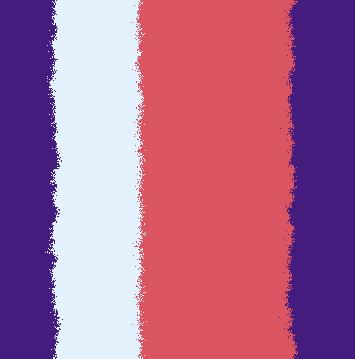}
        \caption{}\label{fig4b}
    \end{subfigure} %
   \begin{subfigure}{.18\textwidth}
        \centering
        \includegraphics[width=32mm]{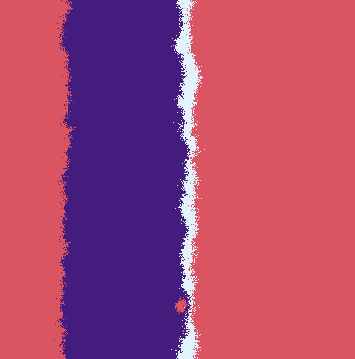}
        \caption{}\label{fig4c}
    \end{subfigure} 
            \begin{subfigure}{.18\textwidth}
        \centering
        \includegraphics[width=32mm]{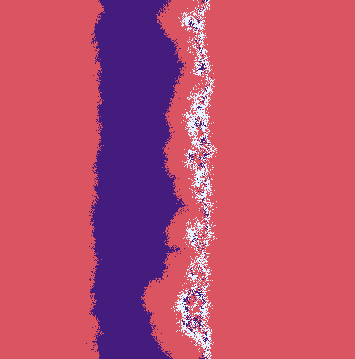}
        \caption{}\label{fig4d}
    \end{subfigure} 
           \begin{subfigure}{.18\textwidth}
        \centering
        \includegraphics[width=32mm]{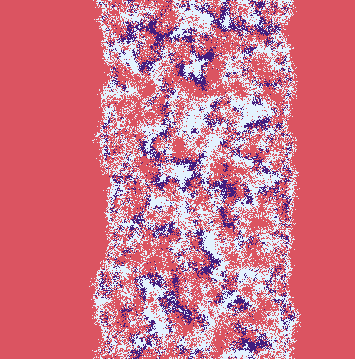}
        \caption{}\label{fig4e}
    \end{subfigure} 
    \caption{Snapshots obtained from a simulation running in lattice with $600^2$ sites starting from the initial conditions in Fig. a, for $R=3$, $\kappa=7.5$, $\alpha=1.0$, $p=m=0.5$. Figs. b, c, d, and e show the spatial configuration after $60$, $192$, $225$, and $435$ generations, respectively.}
  \label{fig4}
\end{figure*}
\begin{figure}
\centering
\includegraphics[width=77mm]{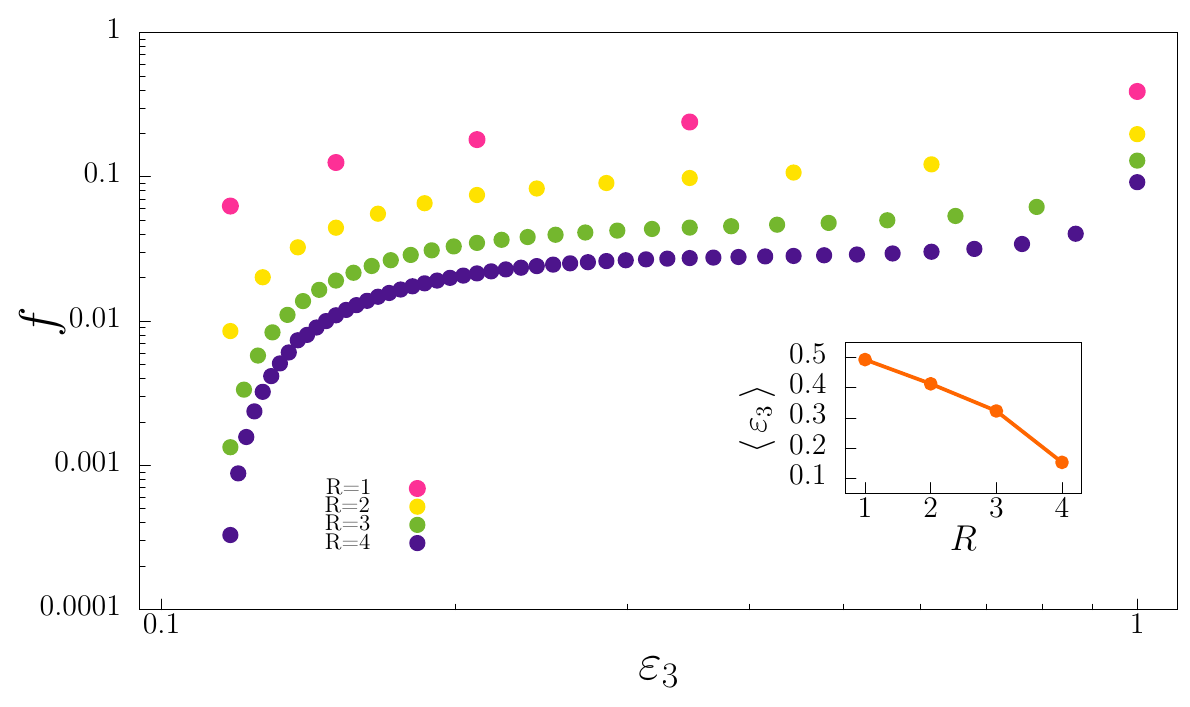}
\caption{Frequency of weakened individuals of species $3$ as a function of predation capacity for various range of antipredator response $R$. The inset shows the mean value of $\varepsilon_3$ for various $R$.}
	\label{fig3c}
\end{figure}
\begin{figure*}
\centering
    \begin{subfigure}{.33\textwidth}
        \centering
        \includegraphics[width=57mm]{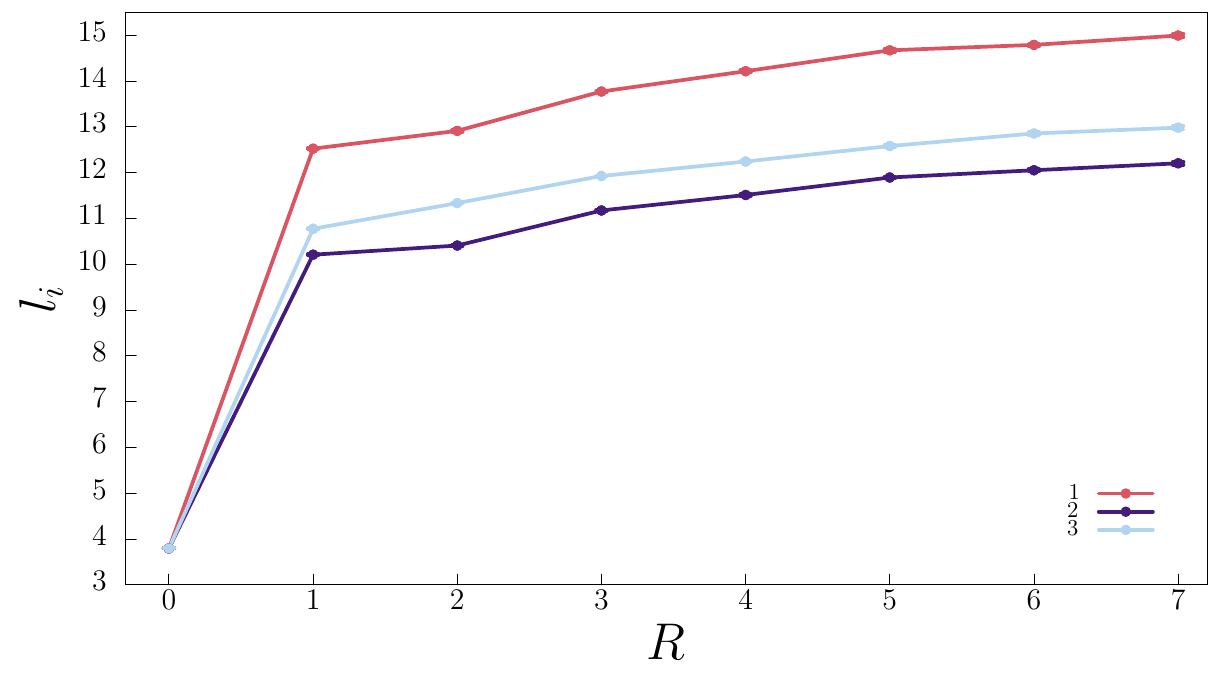}
        \caption{}\label{fig5a}
    \end{subfigure} %
       \begin{subfigure}{.3\textwidth}
        \centering
        \includegraphics[width=57mm]{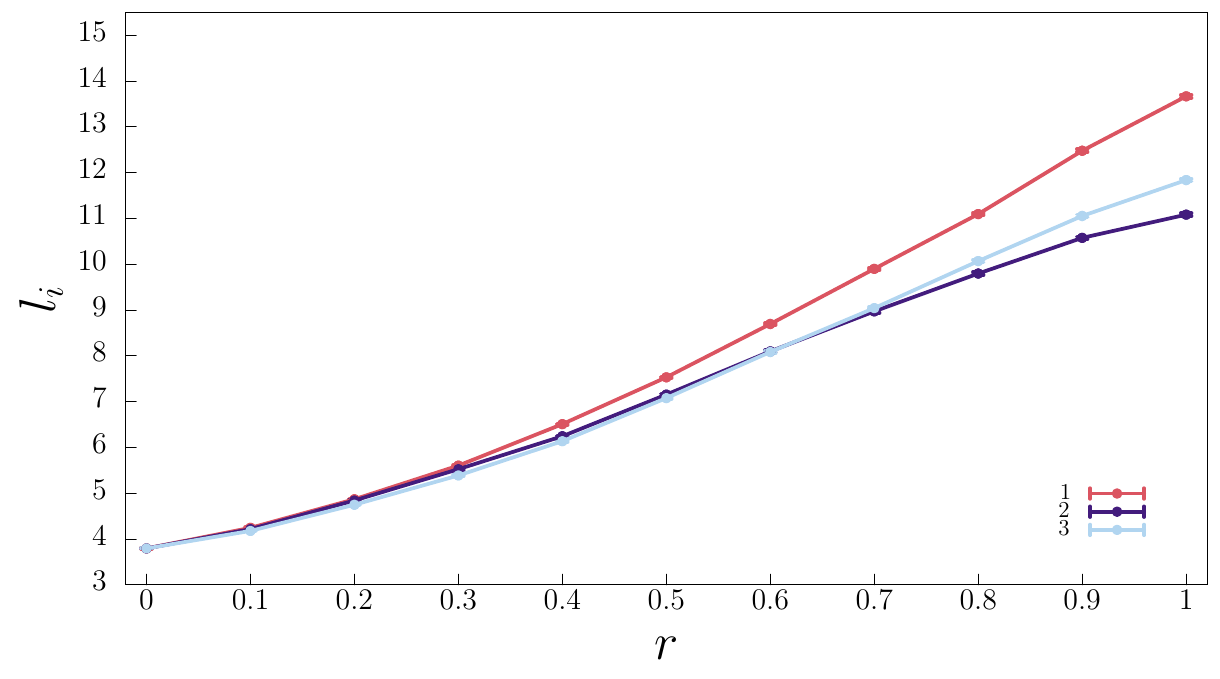}
        \caption{}\label{fig5b}
    \end{subfigure} %
   \begin{subfigure}{.33\textwidth}
        \centering
        \includegraphics[width=57mm]{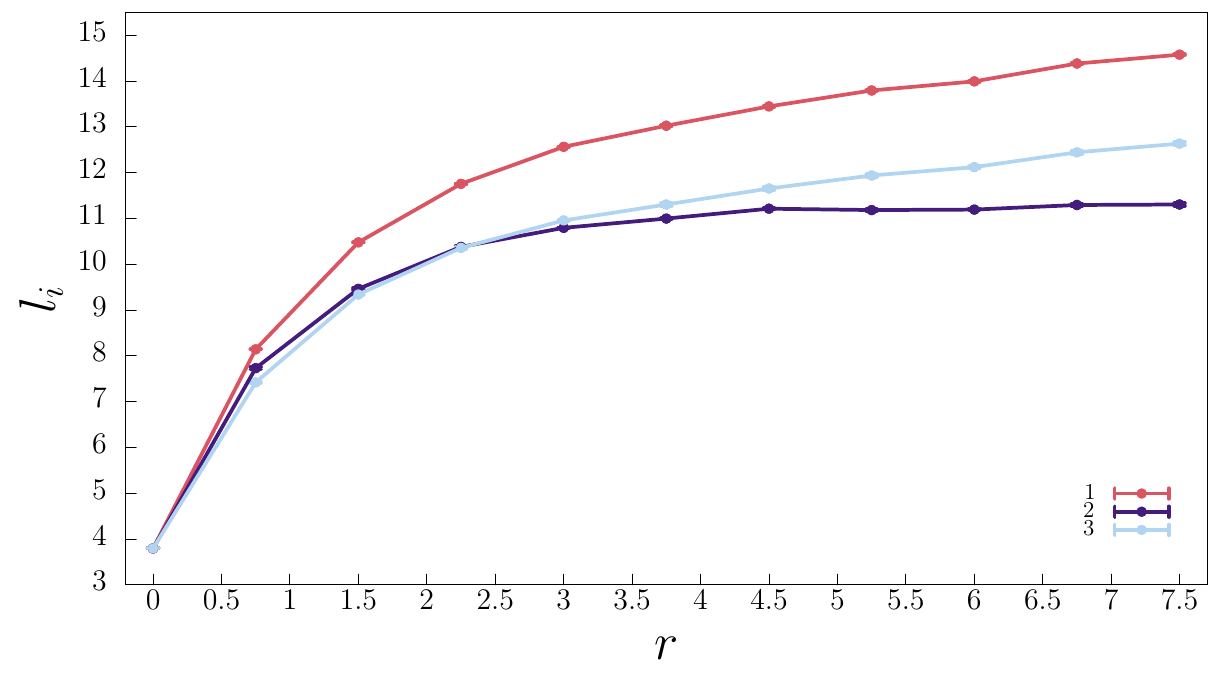}
        \caption{}\label{fig5c}
    \end{subfigure} 
    \caption{Characteristic length of spatial domains in the locally unbalanced rock-paper-scissors model. The results were obtained by analysing the spatial configuration at the end of $100$ simulations in grids with $600^2$ sites running until $5000$ generations.
Figures a, b, and c show the dependence of $l_i$ on the radius of the antipredator response, the conditioning factor, and the antipredator strength factor, respectively. The colours follow the scheme in Fig.~\ref{fig1}.}
  \label{fig5}
\end{figure*}
We now aim to investigate the pattern formation mechanism in more detail. For this purpose, we prepared a single simulation starting from the particular initial condition shown in Fig.~\ref{fig4a}, where each species occupies a third of the grid. The realisation ran in a lattice with $600^2$ sites, for $R=3$, $\kappa=7.5$, $\alpha=1.0$, and $m=p=0.5$. The outcomes are depicted in Fig.~\ref{fig3} and video https://youtu.be/Lsz9E2eENOw; the colours represent the species according to the scheme in Fig.~1. Figures~\ref{fig4b} and ~\ref{fig4c} depict the spatial patterns after $60$, $192$, $225$, and $435$ generations.
As soon as the simulation starts, the rings start moving due to the predator-prey interactions. However, the antipredator response of prey groups of species $1$ hampers the advance of species $3$. According to Fig.~\ref{fig4b}, the consequence is that:
\begin{enumerate}
\item
the red ring enlarges because individuals of species $1$ consume organisms of species $2$ without opposition and defend themselves against predation.
\item
the light blue ring shortens because organisms of species $3$ do not perform the antipredator tactic but suffer resistance of individuals of species $1$;
\item
the  purple ring width is, on average,  constant because organisms do not resist predation or suffer any resistance; the number of
individuals of species 2 consumed (when the red torus surface ring advances over the purple one) is equal, on average, to the number of organisms os organisms of species 3 preyed (when purple ring invades the light blue region).   
\end{enumerate}
To compute the temporal change in each ring width, we consider that the area occupied by species $i$ is defined by the total number of organisms of species $i$. Therefore, 
\begin{equation}
\dot{\delta_i}\,=\,\frac{\dot{I_{i}}}{\sqrt{\mathcal{N}}},
\end{equation}
where $\delta_i$ is the width of the ring occupied by species $i$,
with $i=1,2,3$; the dot stands for the time derivative and $\sqrt{\mathcal{N}}$ is the torus cross section perimeter. 
We calculated the time variation of each ring width for the simulation presented in Fig.~\ref{fig2} using the results for $t\,\leq\,200$ generations, the period that precedes the pattern formation. In this period, the species densities vary linearly in time: $\dot{\delta_1}\,\approx\,0.95$, $\dot{\delta_2}\,\approx\,0$, and $\dot{\delta_3}\,\approx\,-\,0.95$ grid points per generation. 

The narrowing of the ring of species $3$ continues until being so thin that it allows stochastic fluctuations to facilitate the passage of organisms of species $1$ without being caught, as shown in Fig.~\ref{fig4c}. Once individuals of species $1$ reach the purple ring, they multiply because of the abundance of prey. The outcomes show that from the moment organisms of all species meet in the same spatial regions (as in Fig.~\ref{fig2j}), local interactions provokes
the emergence of waves that spread on the entire territory, as one sees in 
Figs.~\ref{fig4d} and \ref{fig4e}. The single-species spatial domains are not symmetric due to the antipredator strategy executed by organisms of species $1$, which imposes species $2$ to propagate in wavefronts shorter than the other species.
\section{The influence of the radius of the antipredator response}
In the previous section, we found that the local antipredator response influences the predation efficiency of species $3$. Now, we aim to find how  the radius of the antipredator response,
$R$,  impacts the average predation capacity reduction. For this reason, we first calculated the frequency of organisms affected differently by the prey opposition for the complete sets of possible prey group sizes. 
The magenta, yellow, green, and purple dots in Fig.~\ref{fig3c} show the frequency of individuals with predation capacity $\varepsilon_3$, for $R=1$, $R=2$, $R=3$, and $R=4$, respectively. We obtained the outcomes running simulations in $300^2$ grid sites with a timespan of $3000$ generations; we assumed the parameters $\kappa=0.75$, $\alpha=1.0$, and $m=p=0.5$.
The inset figure depicts $\langle\,\varepsilon_3\,\rangle$, the mean value of $\varepsilon_3$ during the entire simulation.

According to Eq.~\ref{ht2}, there may be five levels of antipredator response in the case of $R=1$; they are classified according to the group size resisting predation, $g=0,1,2,3,4$. As $R$ increases, the maximum number of prey reacting predator investiture grows: $\mathcal{G}=12$, $\mathcal{G}=28$, and $\mathcal{G}=48$, for $R=2$, $R=3$, and $R=4$, respectively.
 Although species $3$ being, on average, weaker than the others, in terms of predation capacity, the "weakness" is not homogeneously distributed among the organisms. For example, even though that for $R=3$, $\langle\,\varepsilon_3\,\rangle = 0.323\,\langle\,\varepsilon_1\,\rangle =0.323\,\langle\,\varepsilon_2\,\rangle$, this is not a constraining for the majority of the organisms. Figure~\ref{fig3c}  shows  that there is a low frequency of individuals whose predation capacity is  severely decreased  by the antipredator response. 
Moreover, Fig.~\ref{fig3c} reveals that  the effects of the antipredator behaviour increase  if the resistance is less localised. This agrees with a recent publication claiming that the antipredator response is more efficient to reduce the predation risk for a larger $R$ \cite{anti2}.
\section{Characteristic Length Scales}
\begin{figure}[h]
 \centering
       \begin{subfigure}{.4\textwidth}
        \centering
        \includegraphics[width=65mm]{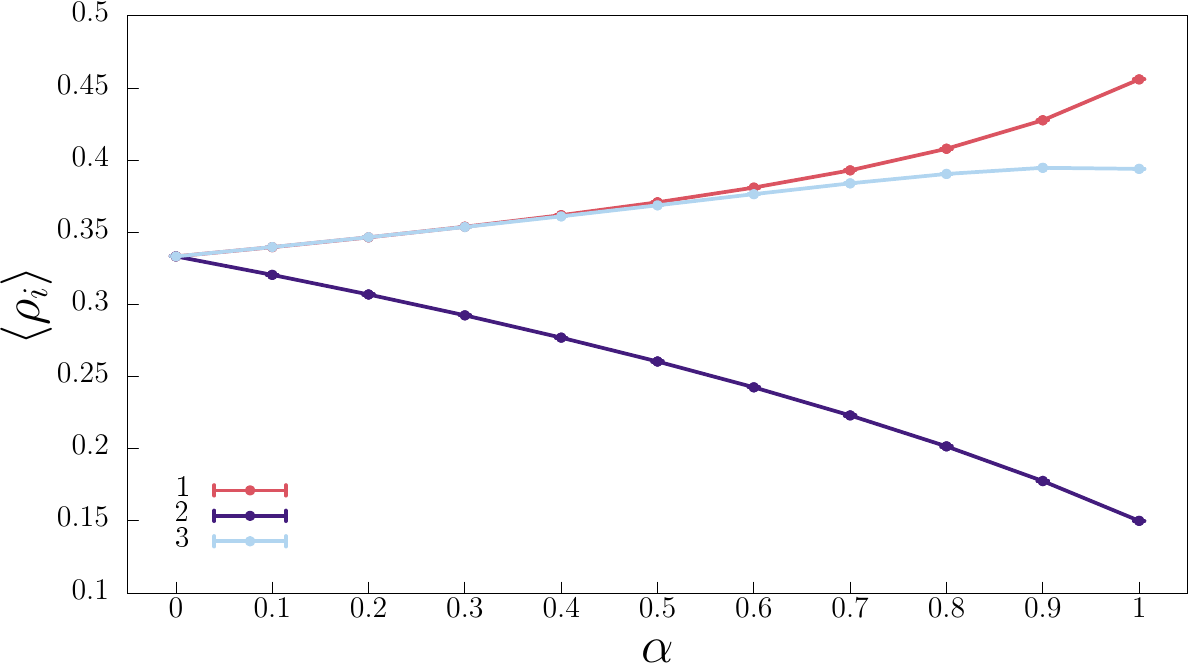}
        \caption{}\label{fig8a}
    \end{subfigure}\\
   \begin{subfigure}{.4\textwidth}
        \centering
        \includegraphics[width=65mm]{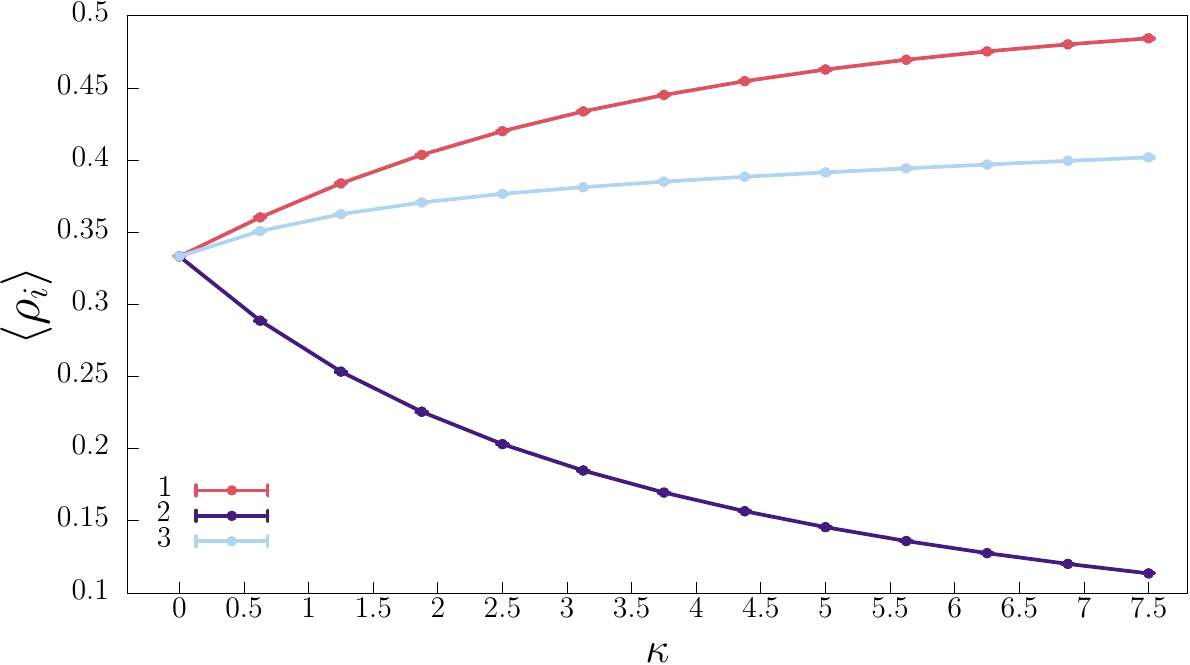}
        \caption{}\label{fig8b}
    \end{subfigure} 
\caption{Mean species densities in terms of the conditioning and antipredator strength factors of species $1$. Figures a and b show the averaged results obtained from the same implementations of the outcomes presented in Figs.\ref{fig5b} and \ref{fig5c}, respectively.}
  \label{fig8}
\end{figure}
The local antipredator reaction of organisms of species $1$ unbalances the spatial rock-papers-scissors game, causing the emergence of domains, inhabited mainly by individuals of the same species.

Now, we aim to calculate the characteristic length which defines the scale of spatial domains occupied by each species.
For this reason, we first calculate the spatial autocorrelation function $C_i(r)$, with $i=1,2,3$, in terms of radial coordinate $r$.

Let us first define the function $\phi_i(\vec{r})$ that represents the presence of an organism of species $i$ in the position $\vec{r}$ in the lattice.
Calculating the mean value $\langle\phi_i\rangle$, we find the Fourier transform
\begin{equation}
\varphi_i(\vec{\kappa}) = \mathcal{F}\,\{\phi_i(\vec{r})-\langle\phi_i\rangle\},
\end{equation}
that is used to compute the spectral densities
\begin{equation}
S_i(\vec{k}) = \sum_{k_x, k_y}\,\varphi_i(\vec{\kappa}).
\end{equation}

The autocorrelation function is given by the normalised inverse Fourier transform
\begin{equation}
C_i(\vec{r}') = \frac{\mathcal{F}^{-1}\{S_i(\vec{k})\}}{C(0)}.
\end{equation}

Finally, we compute the spatial autocorrelation function for species $i$ as 
a function of the radial coordinate $r$:
\begin{equation}
C_i(r') = \sum_{|\vec{r}'|=x+y} \frac{C_i(\vec{r}')}{min\left[2N-(x+y+1), (x+y+1)\right]}.
\end{equation}
The typical size of the spatial agglomerations of organisms of species $i$ is found by 
assuming the threshold $C_i(l_i)=0.15$, where $l_i$ is the characteristic length scale for spatial domains of species $i$ \cite{BAZEIA2022126547,PhysRevE.97.032415}.

We ran a series of $100$ simulations using lattices with $600^2$ grid points starting from different random initial conditions
for $p\,=\,r\,=\,0.5$.
We computed the mean autocorrelation function $C_i(r)$ employing the spatial configuration at $t=5000$ generations. 

First, we investigated how the scale of single-species domains changes with the range of antipredator response $R$. In this set of simulations, we fixed $\kappa=5.0$ and $\alpha=1.0$. 
According to the results depicted in Fig.~\ref{fig5a}, even if the antipredator reaction is limited to $R=1$, the system undergoes pattern formation. Confronted with the standard model, where organisms cannot resist predation ($R=0$), there is an increase in the typical size of the single-species domains, with species $1$ occupying larger territories. The outcomes also show that the less localised the antipredator response (larger $R$), the more extensive are the areas inhabited by individuals of a single species \cite{anti2}. 

Second, considering $R=3$ and $\kappa=5.0$, we studied the dependence of the characteristic length $l_i$ in the percentage 
of individuals of species $1$ conditioned to perform antipredator response. Figure~\ref{fig5b} shows that if no more than $10\%$ of organisms can participate in the antipredator response, the increase in the $l_i$ is approximately the same for every species.
For $0.2\,\leq\,\alpha\,\leq\,0.6$, the outcomes reveal that species $1$ occupy the larger areas of the lattice - followed by species $2$. However, the scenario changes if more than $60\%$ of the organisms of species $1$ are conditioned: agglomerations of species $3$ grow more than clumps of species $2$.

Third, we observed the dependence of the antipredator strength factor in the pattern formation, in the case of all organisms of species $1$ being conditioned, with $R=3$. The outcomes presented in Fig.~\ref{fig5c} show that the typical agglomeration size depends on the intensity of antipredator response, with species $1$ filling the largest single-species domains irrespective of $\kappa$. The second larger agglomerations are formed by organisms of species $2$ in the case of $0\,<\,\kappa\,\leq\,2.25$. In contrast, for $\kappa\,>\,2.25$, $l_2$ is the shortest.
\section{Species Predominance}
We now address the question of whether the local decrease in predation capacity gives the species predominance over the others. 
Figures \ref{fig8a} and \ref{fig8b} depict the mean species densities averaged from the same sets of simulations presented in Fig.~\ref{fig5}. Overall, the outcomes reveal the predominance of species $1$. First, the outcomes show that species $1$ does not predominate only if less than $50\%$ of the organisms are conditioned to participate in the collective strategy; in this scenario, the spatial densities of species $1$ and $3$ are the same, as shown in Fig.~\ref{fig8a}. 
Second, according to Fig.~\ref{fig8b}, for $\alpha=1$,  species $1$ is more abundant  irrespective of the antipredator strength factor. 

Although not being preponderant, our findings also show that the weakening of species $3$ brings positive results in terms of population growth. The outcomes reveal that the stronger is the local opposition faced by the individuals, the higher is the density of species $3$. However, under no circumstances does the weaker species predominate in a locally unbalanced cyclic model; the preponderance is always of the weaker species’ prey.

\begin{figure}
\centering
\includegraphics[width=51mm]{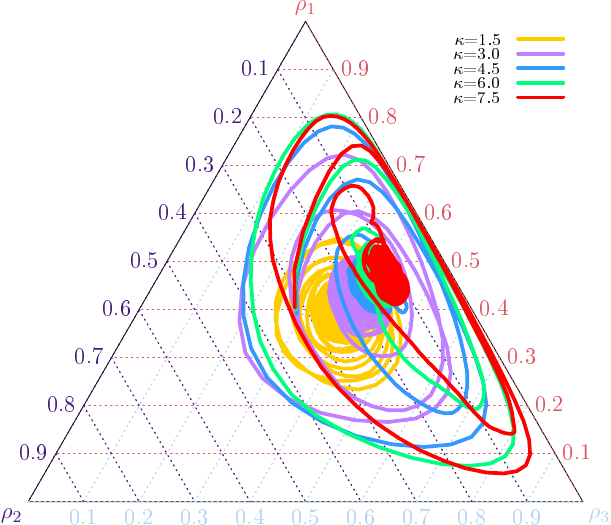}
\caption{Ternary diagram of the species densities for 
various $\kappa$. Each orbit shows $\rho_i$ from a single realisation running in grids with $300^2$ sites, for $R=3$, $\alpha=1.0$, and $p=m=0.5$.}
	\label{fig9}
\end{figure}
\begin{figure}
\centering
\includegraphics[width=68mm]{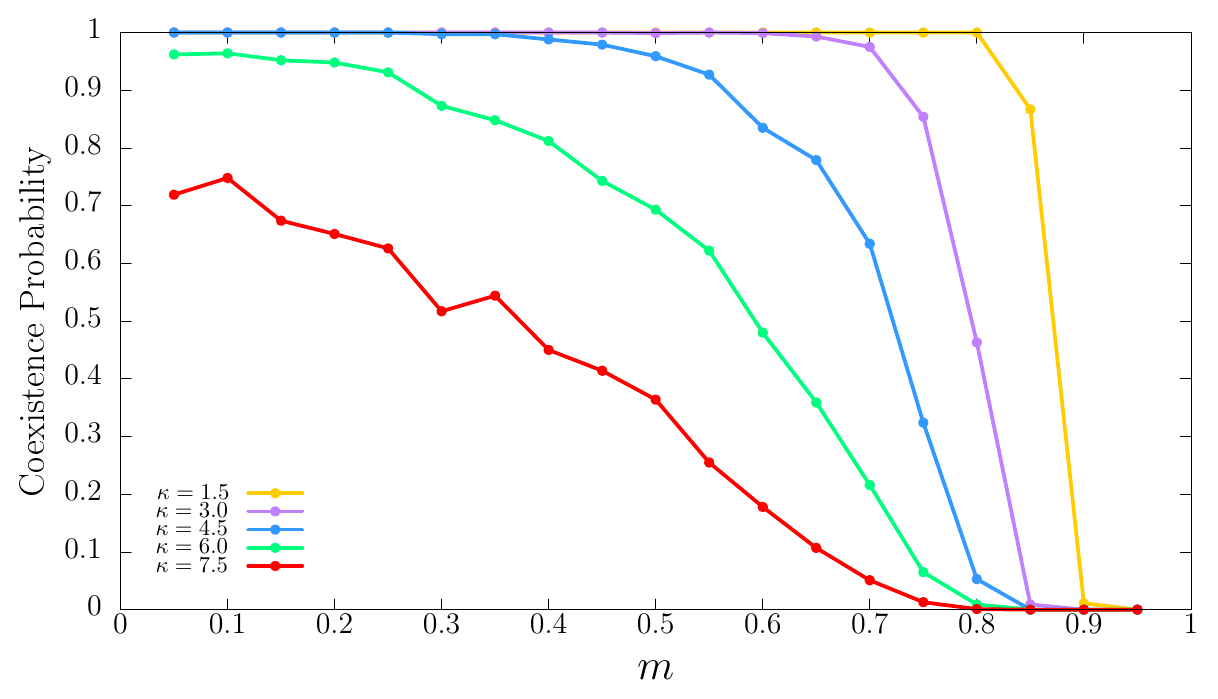}
\caption{Coexistence probability as a function of the mobility probability $m$ for various $\kappa$. The results were obtained by running sets of $1000$ simulations in lattices with $102^2$ grid points, running until $102^2$ generations, assuming $p=1-m$, $R=3$, and $\alpha=1.0$.}
	\label{fig10}
\end{figure}
\section{Coexistence Probability}
Because the antipredator behaviour of organisms of species $1$
unbalances the cyclic spatial rock-paper-scissors model, 
species coexistence may be jeopardised. To investigate this issue, we first observed how the spatial species densities oscillate for various
$\kappa$. The ternary diagram depicted in Figure \ref{fig9} shows the orbits of $\rho_i$ for $\kappa=1.5$ (yellow line), $\kappa=3.0$ (purple line), $\kappa=4.5$ (blue line), $\kappa=6.0$ (green line), and $\kappa=7.5$ (red line). The simulations ran in lattices with $300^2$ sites until $3000$ generations, for $R=3$, $\alpha=1.0$, and $p=m=0.5$. The outcomes show that the species density oscillations in the first stage of the simulations increase with $\kappa$, indicating that the stronger the antipredator reaction of organisms of species $1$, the more the biodiversity may be threatened.

We then investigated the species coexistence as a function of the mobility probability for the cases of Fig.~\ref{fig9}. To this purpose, we implemented different random initial conditions for sets of $1000$ simulations in lattices with $102^2$ grid points for $ 0.05\,<\,m\,<\,0.95$, assuming $R=3$ and $\alpha$=1.0; the predation probability was set to be $p\,=\,1-m$; the simulations ran for a timespan of $102^2$ generations. 
Coexistence occurs if at least one individual of every species is present at the end of the simulation, $I_i (t=5000) \neq 0$ with $i=1,2,3$. This means that if at least one species is absent, the simulation results in extinction. The coexistence probability is defined as the fraction of implementations resulting in coexistence. Figure \ref{fig10} depicts the coexistence probability as a function of $m$ for $\kappa\,=\,1.5$ (yellow line), $\kappa\,=\,3.0$ (purple line), $\kappa\,=\,4.5$ (blue line), $\kappa\,=\,6.0$ (green line), and $\kappa\,=\,7.5$ (red line). 
Overall, biodiversity is threatened because the local antipredator response unbalances the spatial cyclic model. The results show that the effects are highlighted if organisms move with high probability. Moreover, the larger $\kappa$, the more jeopardised the biodiversity is.

\section{Discussion and Conclusions}
\label{sec6}

We investigated the effects of local antipredator response performed by one out of the species in the spatial version of the rock-paper-scissors model. The antipredator reaction is 

initiated
      
whenever a predator tries to consume one of the individuals belonging to the prey group surrounding the predator. The decrease in
predation capacity depends on the prey group size and the antipredator strength of each prey. Besides, participation in the collective reaction depends on the organism's physical and cognitive abilities to properly perform the defence tactic. 

Our findings show that even if a small number of organisms can perform the local behavioural tactic, there are clear benefits for the species that perform the antipredator response. However, the species whose organisms perform the antipredator strategy is not the only one to profit: due to cyclic predator-prey interactions, it is advantageous for the species affected by the local antipredator response. If less than half of individuals are conditioned, both species share territorial dominance. Otherwise, the prevalence is of the species whose organisms behave defensively.

Our results reveal that the local antipredator response unbalances the spatial rock-paper-scissors game differently than if all individuals were equally weakened, independent of the spatial position.
Suppose all organisms of one out of the species are intrinsically weak in terms of effective predation probability, with the frailty being due to the evolutionary characteristics inherent to the species or external interference provoked by any disease or seasonal circumstance. In that case, the reduced predation capacity 
allows its prey to multiply everywhere, reverting in protection against its predator. In those scenarios, the result is a predominance of the weak species \cite{weakest,PedroWeak,uneven}.

Here, the complexity of the local interactions leads to regions with different prey concentrations. In patches with more prey, the local antipredator response is more intense, resulting in a
sharp drop in the organism's predation capacity; thus, the probability of predators invading the prey territory is low. On the contrary, in low prey density regions, a predator is less affected; therefore, the chances of consuming prey is higher, increasing the local predator population. Besides organisms with low and high predation capacity being concentrated in different patches, our results show that
the number of organisms with low predation capacity is much smaller.
If the fraction of organisms conditioned to perform the antipredator strategy is small, the total areas occupied by prey and predators
are approximately the same. However, if the fraction of conditioned species is greater than $50\%$, the fraction of territory occupied by predators grows faster than the prey-dominated areas. Therefore, the weaker species does not dominate under any circumstances if the antipredator response locally causes the weakening. 

Despite the indisputable benefits of the collective antipredator behavioural strategy, biodiversity may be jeopardised if the antipredator resistance is too strong. This happens because of the species' densities oscillations in the transient pattern formation stage. As expected, the chances of the local defensive strategy affecting coexistence accentuate if individuals move with higher mobility probabilities. 
Our outcomes may also be helpful to ecologists to model biological systems where the consequences of behavioural strategies in the local interactions play a vital role in biodiversity conservation.
\section*{Acknowledgments}
We thank CNPq, ECT, Fapern, and IBED for technical support.

\nocite{*}
\bibliography{LocalAntipredatorRPS}

\end{document}